\documentclass[aps,prl,reprint,amsmath,amssymb,superscriptaddress]{revtex4-2}

\usepackage{graphicx}
\usepackage{dcolumn}
\usepackage{bm}
\usepackage{color}
\usepackage{subfigure}
\usepackage{epsf}
\usepackage{amsmath,amstext,amssymb}
\usepackage{enumitem}
\usepackage[colorlinks=true,citecolor=MidnightBlue,linkcolor=MidnightBlue,urlcolor=MidnightBlue]{hyperref}
\usepackage[ansinew]{inputenc}
\usepackage{braket}
\usepackage[usenames,dvipsnames]{xcolor}
\usepackage{bbold}
\usepackage{siunitx}
\PassOptionsToPackage{numbers,sort&compress}{natbib}

\graphicspath{{figures/}}
\makeatletter
\g@addto@macro\normalsize{%
	\setlength\abovedisplayskip{4pt}
	\setlength\belowdisplayskip{4pt}
	\setlength\abovedisplayshortskip{4pt}
	\setlength\belowdisplayshortskip{4pt}
}
\makeatother

\begin{document}

\title{Detecting heat leaks with trapped ion qubits}

	\author{D.~Pijn}
	\affiliation{Institut f\"ur Physik, Universit\"at Mainz, Staudingerweg 7, 55128
		Mainz, Germany}
	\author{O.~Onishchenko}
	\affiliation{Institut f\"ur Physik, Universit\"at Mainz, Staudingerweg 7, 55128
		Mainz, Germany}
	\author{J.~Hilder}
	\affiliation{Institut f\"ur Physik, Universit\"at Mainz, Staudingerweg 7, 55128
		Mainz, Germany}
	\author{U.~G.~Poschinger}\email{poschin@uni-mainz.de}
	\affiliation{Institut f\"ur Physik, Universit\"at Mainz, Staudingerweg 7, 55128
		Mainz, Germany}
	\author{F.~Schmidt-Kaler}
	\affiliation{Institut f\"ur Physik, Universit\"at Mainz, Staudingerweg 7, 55128
		Mainz, Germany}
	\author{R.~Uzdin}
	\affiliation{Fritz Haber Research Center for Molecular Dynamics,Institute of Chemistry, The Hebrew University of Jerusalem, Jerusalem 9190401, Israel}

\date{\today}

\begin{abstract}
	Recently, the principle of \textit{passivity} has been used to set bounds on the evolution of a microscopic quantum system with a thermal initial state. In this work,  we experimentally demonstrate the utility of two passivity based frameworks: global passivity and passivity deformation, for the detection of a ``hidden'' or unaccounted environment. We employ two trapped-ion qubits undergoing unitary evolution, which may optionally be coupled to an unobserved environment qubit. Evaluating the measurement data from the system qubits, we show that global passivity can verify the presence of a coupling to an unobserved environment - a heat leak - in a case where the second law of thermodynamics fails. We also show that passivity deformation is even more sensitive, detecting a heat leak where global passivity fails.
 \end{abstract}

\maketitle

\textit{Introduction - } Thermodynamics has originally been conceived as a practical theory for describing heat flows and efficiencies of heat engines. Taking the point of view that quantum devices emerge to be the ``heat engines'' of the 21$^{st}$ century motivates to explore if and how quantum thermodynamics can contribute to the maturation of quantum technologies. In stochastic thermodynamics, fluctuation theorems \cite{Seifert2007FTRev,FThanggi} and thermodynamic uncertainty relations have been formulated \cite{TUR1,TUR2}, and in quantum thermodynamics, advanced master equations, resource theory \cite{RT1,RT2}, passivity-based frameworks \cite{Uzdin2018,Uzdin2021}, and entropy-based methods \cite{StrasbergObserv,bera2019thermodynamics} have been developed. Recent years have seen increasing experimental studies on demonstrations and validations \cite{Exp3ions,ExpHuardDemon,ExpMainzFlywheel,ExpMeasEngine,ExpMurch,ExpPekola,ExpPoem,ExpRoss,ExpSerra,ExpCampisi2020DWave,ExpDeffner2018ErrorAnnealers,ExphenaoIBM,ExpCampisiFT,henao2021experimental}. 

\begin{figure}[ht!p]
    \centering
    \includegraphics[width=\columnwidth,trim={2.7cm 7.5cm 2.5cm 7.5cm},clip]{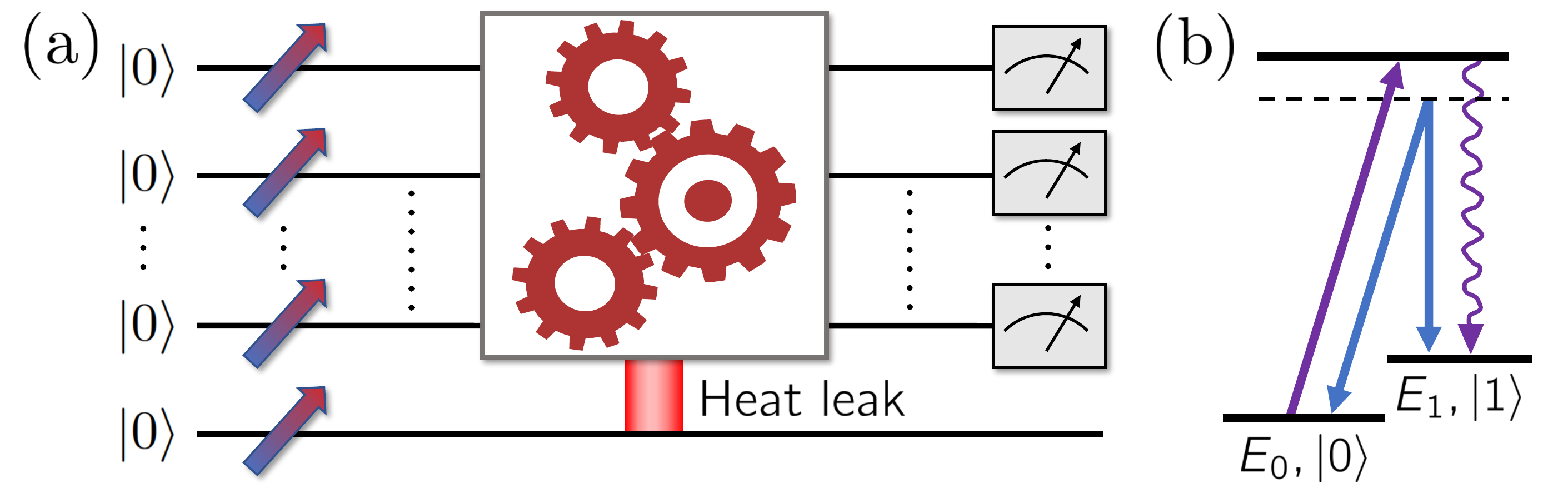}
    \caption{\textbf{(a)} A number of system qubits and one environment qubit are initialized to thermal states. The system qubits undergo a unitary ``black box'' evolution and are measured. In the case of a heat leak, the system is coupled to the undetected environment qubit. \textbf{(b)} Laser-driven operations transfer the qubit states between $\ket{0}$ and $\ket{1}$, with respective energy eigenvalues.}
    \label{fig:intro}
\end{figure}

Quantum information processing (QIP) devices, such as quantum computers and  quantum simulators, suffer from the inherent fragility of quantum states with respect to environmental perturbations. This renders QIP devices to be susceptible to various error mechanisms. In this work, we show that thermodynamics-inspired frameworks are relevant for the characterization of the operation of presently available well-controlled quantum systems. We use a trapped-ion setup to experimentally study two recently developed passivity-based frameworks: \textit{global passivity} \cite{Uzdin2018}, and \textit{passivity deformation} \cite{Uzdin2021}.

A wide variety of methods has been developed for benchmarking QIP devices and their operational building blocks, ranging from low-level techniques such as quantum process tomography \cite{Haeffner2005} to holistic high-level approaches such as quantum volume measurement \cite{Cross2019,Pino2021}. Methods developed within microscopic thermodynamics offer complementary approaches for characterizing the performance of QIP devices. Thermodynamic constraints such as the microscopic second law \cite{Raam2ndLawReview,Esposito2011EPL2Law,PeresBook} set constraints on the allowed dynamics of mixed states in an isolated system. A violation of these constraints provides information on undesired interaction with an external environment. Crucially, these constraints make no assumptions on the tested protocol (a ``black box'' test), and are therefore agnostic to the complexity of the evolution. Moreover, thermodynamics-based tests are insensitive to coherent errors that arise due to miscalibration. This represents a useful feature in view of the identification and mitigation of error sources.

Not all thermodynamic constraints are scalable in the sense of providing realistic measurement protocols for increasing system  sizes. The analogue of Clausius inequality in microscopic systems \cite{Raam2ndLawReview,Esposito2011EPL2Law,PeresBook} requires quantum state tomography for evaluating changes in the von Neumann entropy. The measurement of trajectories in fluctuation theorems \cite{Seifert2007FTRev,FThanggi} is equivalent to  classical process tomography, and resource theory \cite{RT1,RT2} also requires state tomography for evaluating the R{\'e}nyi divergence. \\
Passivity-based bounds, based on expectation values of observables, provide polynomial scaling of the number of measurements with respect to the system size \cite{ExphenaoIBM}.
The potential practical use of a thermodynamic bound is also determined by its tightness. For example, \textit{global passivity} \cite{Uzdin2018} and the second law set intrinsically loose bounds when the thermal environment is small, therefore their accuracy and predictive power are inherently limited. However, \textit{passivity deformation} \cite{Uzdin2021} provides increased sensitivity, since the constraints are tight by construction, independently of the sizes of the system constituents. 

In this work, a trapped-ion based quantum computer is used to demonstrate that global passivity can be more sensitive to a heat leak - a spurious energy exchange channel to environmental degrees of freedom - as compared to the second law, and that passivity deformation is more sensitive than global passivity. To achieve this, we use an unobserved thermal qubit as a controllable environment. Two system qubits constitute the ``visible'' part of the system, i.e. qubits employed for the execution of a QIP protocol. The goal is to detect the interaction with the environment qubit by measuring only the system qubits.

\textit{Theory - } We consider a system consisting of degrees of freedom on which measurements can be performed. If the system is isolated in the sense that only classical (possibly noisy) external driving fields are applied, an initial state $\hat{\rho}_0$ of the system will evolve into a final state
\begin{equation}
    \hat{\rho}_f=\sum_k p_k \hat{U}_k\hat{\rho}_0\hat{U}_k^{\dagger}.
    \label{eq:mixUnitaries}
\end{equation}
This evolution is determined by a mixture of unitary transforms $\hat{U}_k$. As such evolutions are \textit{unital} (i.e. a fully mixed state is invariant under such transformation),  the entropy of the system will always increase \cite{mendl2009unital}. A unital evolution can be interpreted as the result of a classical noisy driving field.  Quantum evolution that cannot be written as (\ref{eq:mixUnitaries}) ultimately requires an interaction of the system with some external environment, e.g. an ancilla or a thermal bath. Therefore, a verification that the evolution is not of the form (\ref{eq:mixUnitaries}) via observations on the system confirms the presence of a \textit{heat leak}.\\
The frameworks employed in this work for the detection of heat leaks rely on the notion of passivity. In general, an operator $\hat{A}$ is \textit{passive} with respect to another operator $\hat{B}$, if $[\hat{A},\hat{B}]=0$, i.e. a common set of eigenvectors exists, and if decreasingly ordered eigenvalues of $\hat{A}$ correspond to increasingly ordered eigenvalues of $\hat{B}$. For example, if a density operator $\hat{\rho}$ is passive with respect to the system Hamiltonian $\hat{H}$, eigenstates are less populated for increasing energy eigenvalues. The most common example is a thermal (Gibbs) state expressed in the energy eigenbasis, where the occupation probabilities monotonically decrease with the energy eigenvalue.  Physically, this has the important consequence that no energy can be extracted from a passive state via unitary coupling to an external work body \cite{ALLAHVERDYAN2004}. In the next section, we briefly outline the \textit{global passivity} and \textit{passivity deformation} frameworks.

\textit{Global passivity - } The global passivity inequalities impose bounds for changes of expectation values of a certain class of observables. We consider any unital process (Eq. \ref{eq:mixUnitaries}) taking initial state $\hat{\rho}_0$ to the final state $\hat{\rho}_f$, and a function $F(x)$ which is monotonically decreasing on $\min [\text{eig}(\hat{\rho}_0)]\le x\le\max [\text{eig}(\hat{\rho}_0)]$. By construction, $F(\hat{\rho}_0)$ is passive with respect to $\hat{\rho}_0$, and the global passivity inequalities assume the form:
\begin{eqnarray}
\delta \langle F(\hat{\rho}_0)\rangle &=& \langle F(\hat{\rho}_0)\rangle_{f}-\langle F(\hat{\rho}_0)\rangle_{0} \nonumber \\
&=&\text{tr}[F(\hat{\rho}_{0})(\hat{\rho}_{f}-\hat{\rho}_{0})]\ge 0
\label{eq: F GP}
\end{eqnarray}
Any violation of this inequality is a sufficient condition for the evolution $\hat{\rho}_0\rightarrow \hat{\rho}_f$ to be not of the form (\ref{eq:mixUnitaries}) and therefore indicates the presence of a heat leak.

Equation \ref{eq: F GP} leads to a simple form of the second law in a microscopic setup. We consider a setup comprised of a cold object $c$ and a hot object $h$, each described by Hamiltonian $\hat{H}_{c,h}$, and initial thermal states
\begin{equation}
\hat{\rho}_{j}^{(0)}=e^{-\beta_{j}\hat{H}_{j}}/Z_j \qquad j=c,h
\label{eq:thermalstates}
\end{equation}
where $Z_j=\text{tr}[e^{-\beta_{j}\hat{H}_{j}}]$. The initial state of the joint cold/hot system is uncorrelated
\begin{equation}
\hat{\rho}_{0}=\hat{\rho}_{c}^{(0)}\otimes\hat{\rho}_{h}^{(0)}=\frac{e^{-\beta_{c}\hat{H}_{c}\otimes \mathbb{1}_{h}-\beta_h \mathbb{1}_{c}\otimes \hat{H}_{h}}}{Z_{c}Z_{h}}
\end{equation}

Setting $F(x)=-\ln x $ yields
\begin{equation}
F(\hat{\rho}_{0})=\beta_{c}H_{c}\otimes \mathbb{1}_{h}+\beta_h \mathbb{1}_{c}\otimes H_{h}-\ln(Z_{c}Z_{h})\mathbb{1}_{ch}
\end{equation}

where $\mathbb{1}_{ch}$ is the identity operator on the product Hilbert space of systems $c$ and $h$. The term $-\ln(Z_{c}Z_{h})\mathbb{1}_{ch}$ ensures positivity of $F(\hat{\rho}_0)$. However, upon taking the difference  $\langle F(\hat{\rho}_0)\rangle_{f}-\langle F(\hat{\rho}_0)\rangle_{0}$,  this term cancels out.

Now Eq. \ref{eq: F GP} reads
\begin{equation}
\beta_{c}\;\delta\left\langle \hat{H}_{c}\right\rangle +\beta_{h}\;\delta\left\langle \hat{H}_{h}\right\rangle \ge 0
\label{eq:gp1ineq}
\end{equation}
which is one possible form of the second law for microscopic systems, i.e. the analogue of the classical Clausius theorem $\oint\frac{\delta Q}{T}\geq 0$. 
 
We obtain a more general, parametric set of inequalities by using $F(x)=\text{sgn}(\alpha)(-\ln x)^{\alpha}$, which is a monotonically increasing function on $0\le x\le1$. We introduce
\begin{equation}
    \hat{B}=\beta_{c}\hat{H}_{c}+\beta_{h}\hat{H}_{h}-d\;\mathbb{1}_{ch}
\end{equation}
and the shorthand notation
\begin{equation}
    F(\hat{B})=\hat{B}^{\alpha}=\text{sgn}(\alpha)(\beta_{c}\hat{H}_{c}+\beta_{h}\hat{H}_{h}-d_{\epsilon}\mathbb{1}_{ch})^{\alpha}
\label{eq:GPBtoalpha}
\end{equation}
where the choice 
\begin{equation}
d =\min(\text{eig}(\beta_{c}\hat{H}_{c}+\beta_{h}\hat{H}_{h}))+\epsilon
\end{equation}
with a sufficiently small $\epsilon$ enforces nonzero and positive eigenvalues of $\hat{B}$. Invoking Eq. \ref{eq: F GP}, we obtain the global passivity inequalities in a compact form:
\begin{equation}
\delta\langle \hat{B}^{\alpha}\rangle \geq 0.
\label{eq:gp3ineq}
\end{equation}
Note that the microscopic form  of the second law, Eq. \ref{eq:gp1ineq}, is obtained from Eq. \ref{eq:gp3ineq} for $\alpha=1$. A violation of the inequality Eq. \ref{eq:gp3ineq} necessarily implies that one of the underlying assertions are violated, which is either that the initial state is not passive or that the system evolution is not of the form (\ref{eq:mixUnitaries}) and therefore an interaction with environmental degrees of freedom is present.

For $\alpha \gg 1$, the largest eigenvalue
of $\hat{B}$ dominates, since the it corresponds
to the smallest eigenvalue of $\hat{\rho}_{0}$, as $\hat{B}$ is obtained
from $\hat{\rho}_{0}$ using a monotonically decreasing function. Thus,
as $\alpha\to+\infty$, inequality Eq. \ref{eq:gp3ineq} states that the probability of the state that was initially least populated cannot decrease beyond its initial value. Conversely, for $\alpha\to-\infty$ we learn that the probability of the state that was initially most populated cannot
increase beyond its initial value. Different values of $\alpha$ put emphasis on different eigenvalues
of the expectation values and therefore can potentially detect different
types of heat leaks.

\textit{Passivity deformation - } Passivity deformation \cite{Uzdin2021} is a general and versatile framework for deriving passivity-based inequalities. It follows from the observation that a globally passive operator $\hat{B}$ can be used to generate inequalities involving an observable $\hat{A}$. We introduce the operator
\begin{equation}
\hat{B}'=\hat{B}+\xi \hat{A}
\label{eq:PD0}
\end{equation}
where $\xi$ is a real deformation parameter and $\hat{A}$ is an observable of interest that satisfies $[\hat{B},\hat{A}]=0$. Hence, $\hat{B}$ and $\hat{B}'$ have the same eigenstates. If, moreover, the eigenvalue ordering of $\hat{A}$ and $\hat{B}$ is the same, the global passivity of $\hat{B}$ is inherited by $\hat{B}'$ and the inequality 
\begin{equation}
\delta\langle \hat{B}'\rangle \geq 0
\label{eq:PD1}
\end{equation}
holds for any unitary evolution. This condition is trivially satisfied for $\xi=0$, while it is violated for large deformations, since $\hat{A}$ is in general not globally passive. Thus, there exist extremal values $\xi_m\le0$ and $\xi_p\ge0$ such that Eq. \ref{eq:PD1} holds for $\xi_m < \xi < \xi_p$. If the eigenvalues of $\hat{B}$ and $\hat{A}$ are known, finding the limit values $\xi_{m,p}$ analytically or numerically is a simple task. The obtained passivity deformation inequalities
\begin{equation}
\delta\langle\hat{A}\rangle  \geq  -\frac{1}{\xi} \delta\langle \hat{B}\rangle 
\qquad \forall \qquad \xi_m \leq \xi \leq \xi_p
\label{eq:PD2}
\end{equation}
 First, they may describe observables beyond global energetics (e.g. the ground state population of a sub-system). Furthermore, unlike global passivity, the passivity deformation inequalities Eqs. \ref{eq:PD2} are guaranteed to be tight for some nontrivial process $\hat{\rho}_0\rightarrow \hat{\rho}_f$, even when the thermal environment is small. Consequently, as demonstrated in our experiment, the passivity deformation inequalities may have stronger sensitivity to violation of unitality.

\begin{figure}[h]
    \centering
    \includegraphics[width=\columnwidth,trim={1.5cm 2.5cm 1.5cm 2.5cm},clip]{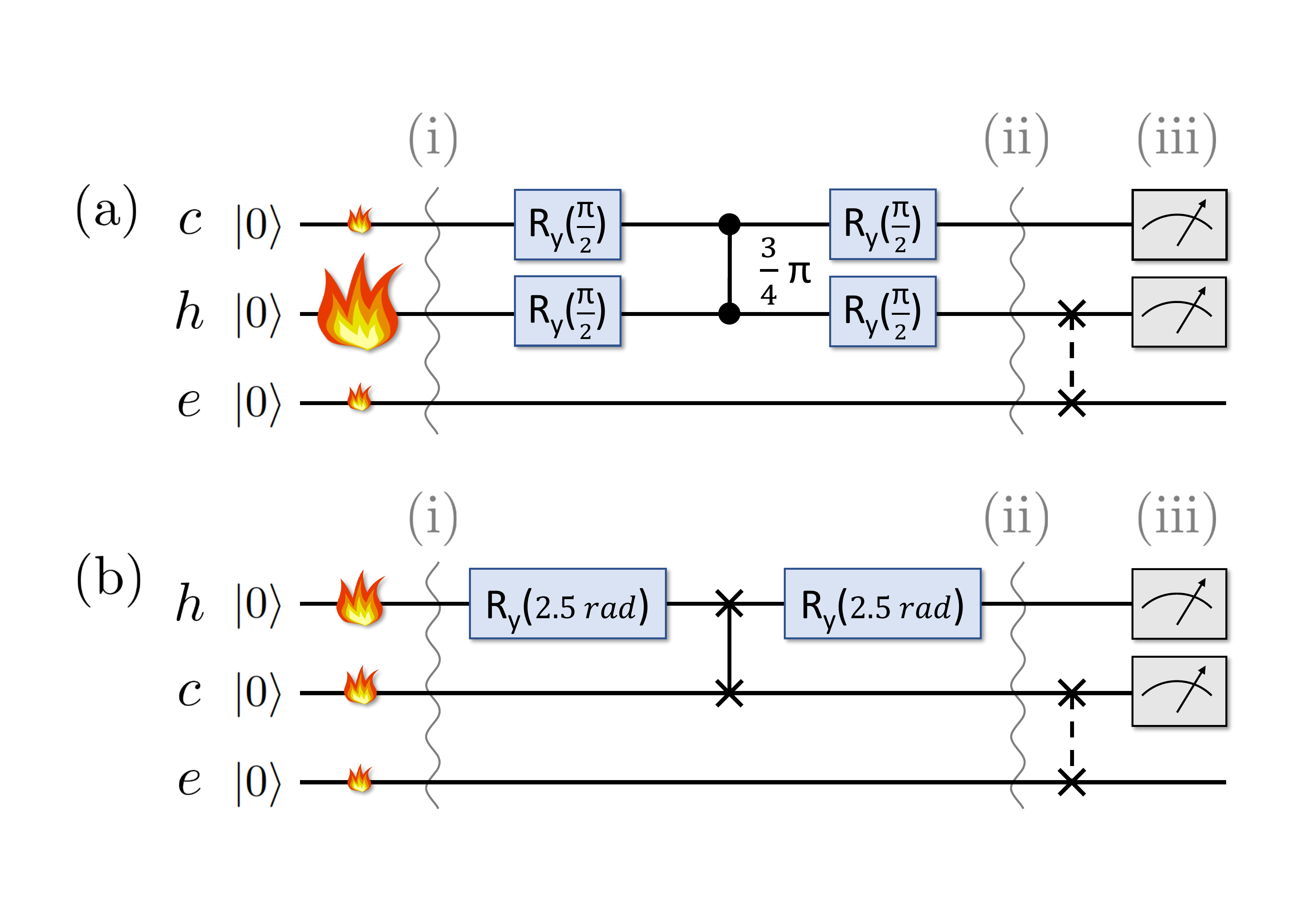}
    \caption{Quantum circuits used for demonstrating a violation of \textbf{(a)} 
    the global passivity inequality (Eq. \ref{eq:gp1ineq}), and \textbf{(b)} the passivity deformation inequality (Eq. \ref{eq:PD2}). In both cases, all qubits are initialized to thermal states (before \textit{(i)}), the system qubits undergo unitary evolution (before \textit{(ii)}), and one of the system qubits is optionally swapped with the environment qubit before measurement \textit{(iii)}. Generation of thermal (mixed) states is depicted by a flame, with different spin temperatures proportional to the size of the flame. The temperatures and the particular unitary quantum operations before the final SWAP gate are chosen to provide optimum sensitivity for the detection of the heat leak.}
    \label{fig:circuits}
\end{figure}

\textit{Experiment - } The platform we employ in this work for showing the violation of passivity-based inequalities is based on qubits encoded in trapped atomic ions. The ions are confined in a microstructured, segmented radio frequency trap \cite{Kaushal2020}, and can be moved between different storage sites via shuttling operations \cite{KIELPINSKI2002,WALTHER2012}. Laser beams, directed to a fixed storage site - the laser interaction zone (LIZ), are employed for initialization, manipulation and readout of the qubits. This way, any single qubit or a pair of qubits can undergo a laser-driven operation in the LIZ, without any crosstalk affecting the remaining qubits stored at different trap sites. The qubits are encoded in the spin degree of freedom of the valence electron of $^{40}$Ca$^+$ ions \cite{POSCHINGER2009}, i.e. the qubit states correspond to the electronic states $\ket{0}\equiv \ket{S_{1/2},m_J=-1/2}$ and $\ket{1}\equiv \ket{S_{1/2},m_J=+1/2}$. The qubits states feature an energy splitting of $\omega_0 \approx 2\pi\times$10~MHz via the Zeeman effect caused by a static externally applied magnetic field \cite{RusterLongLived2016}. However, without loss of generality, we assign the dimensionless energy eigenvalues $E_0=0$ and $E_1=1$ in the following. The free Hamiltonian for both qubits thus reads
\begin{equation}
    \hat{H}^{(0)}_j=\ket{1_j}\bra{1_j} \qquad j=c,h
    \label{eq:freehamil}
\end{equation}
The qubits are read out via laser-driven, selective population transfer to the metastable $D_{5/2}$ state, followed by detection of state-dependent laser-induced fluorescence \cite{POSCHINGER2009}. This way, above-threshold detection of fluorescence corresponds to the qubit being detected in $\ket{0}$, while below-threshold detection corresponds to the qubit being detected in $\ket{1}$. Repeated execution of a given protocol therefore yields estimates of the occupation probabilities for each logical basis state of the qubit register. The relevant error sources are given by shot noise for a finite number of shots, yielding statistical errors, and state preparation and measurement (SPAM) errors, leading to systematic errors. \\
Our experimental protocols employ a `cold' qubit $c$, a `hot' qubit $h$ and a third, unobserved environment qubit $e$. At the beginning of each experimental sequence, these are successively initialized to thermal states Eq. \ref{eq:thermalstates} with respect to the free Hamiltonian Eq. \ref{eq:freehamil} via incomplete optical pumping \cite{PhysRevLett.123.080602}. Here, any desired spin temperature can be preset via control of the pump laser pulse duration, such that the inverse temperatures $\beta_j$ for $j=c,h,e$ (in terms of the dimensionless energy eigenvalues) are given from the Boltzmann weights via
\begin{equation}
\beta_j=\ln\left(\frac{p_0^{(j)}}{1-p_0^{(j)}}\right),
\end{equation}
where $p_0^{(j)}$ is the population of state $\ket{0_j}$. \\

\begin{figure}[h]
    \centering
    \includegraphics[width=0.9\columnwidth]{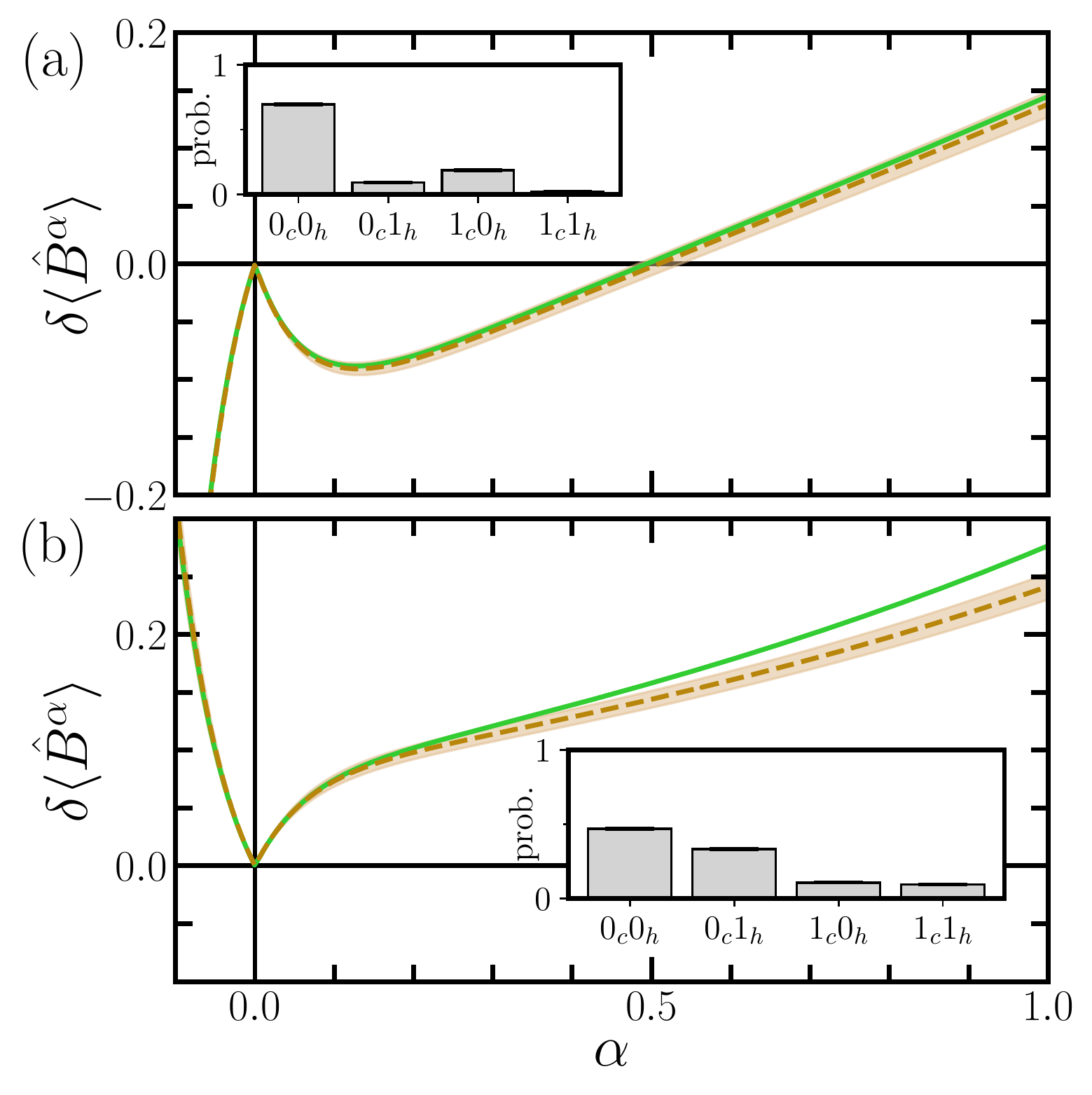}
    \caption{For the first protocol, depicted in Fig. \ref{fig:circuits}(a), $\delta\langle \hat{B}^{\alpha}\rangle$ is shown as a function of $\alpha$. In \textbf{(a)}, the system qubits are coupled to the environment qubit via the final SWAP gate, and we observe $\delta\langle \hat{B}^{\alpha}\rangle \leq 0$ for $\alpha \lesssim$~0.5. In \textbf{(b)}, the SWAP gate is not performed, and we observe no violation of the global passivity inequality (Eq. \ref{eq:gp3ineq}). The experimental data (brown, dotted) is compared to the theoretical expectation values (green). The inset shows the measured binary probabilities for system qubits $c,h$ from which the experimental data was calculated. 6700 shots were used.}
    \label{fig:result1}
\end{figure}

\textit{Heat leak detection via global passivity - }  Qubits $c$, $h$ and $e$ are successively moved to the LIZ and initialized to thermal state. The inverse temperatures $\beta_c =\ $2.23(4), $\beta_h =\ $0.43(2) and $\beta_e =\ $2.02(4) are chosen to provide optimum sensitivity for the detection of the heat leak.\\
After initialization, the system qubits $c$ and $h$ are stored pairwise at the LIZ and undergo a laser-driven unitary evolution. For our protocol, this evolution consists of a two-qubit Ising-type phase gate mediated by light-shifts \cite{LEIBFRIED2003A}, described by 
\begin{eqnarray}
\{\ket{0_c0_h},\ket{1_c1_h}\} &\rightarrow& e^{i\Phi}\{\ket{0_c0_h},\ket{1_c1_h}\} \nonumber \\
\{\ket{0_c1_h},\ket{1_c0_h}\} &\rightarrow& \{\ket{0_c1_h},\ket{1_c0_h}\}. 
\end{eqnarray}
We chose $\Phi=3\pi/4$ to provide optimum sensitivity to the heat leak. This gate is sandwiched between two local qubit rotations by angle $\pi/2$:
\begin{equation}
    \hat{U}_y = \exp\left(-i\frac{\pi}{4}(\hat{\sigma}_y^{(c)}\oplus  \hat{\sigma}_y^{(h)})\right),
\end{equation}
where $\hat{\sigma}_y^{(j)}$ is the Pauli $Y$ operator for qubit $j$. The quantum circuit for this protocol is depicted in Fig. \ref{fig:circuits}(a). After the coherent evolution, qubits $c$ and $h$ are separated \cite{PhysRevA.90.033410}, then qubits $h$ and $e$ are merged to the LIZ, where they can undergo an optional SWAP gate. The SWAP gate is executed via physical swapping of the ion positions, which has been shown to realize a unit-fidelity gate in \cite{PhysRevA.95.052319}, as the ions are indistinguishable and the control over the operations exerted via electric fields, which do not affect the qubit. Finally, \textit{only} the system qubits $c$ and $h$ are read out as described above.\\
Each single shot $k$ yields one of the results $\{\ket{0_c0_h},\ket{0_c1_h},\ket{1_c0_h},\ket{1_c1_h}\}$, corresponding to the measured energies $E^{(k)}_j=\{0,+1\}$. This yields the single-shot measurement result of operator $\hat{B}^{\alpha}$ Eq. \ref{eq:GPBtoalpha} via
\begin{equation}
B_{(k)}^{\alpha}=\text{sgn}(\alpha)(\beta_{c}E_{c}^{(k)}+\beta_{h}E_{h}^{(k)}-d_{\epsilon})^{\alpha}.
\label{eq:singleshotresult}
\end{equation}
From $N$ acquired shots, we evaluate the expectation values by evaluating sample averages based on the obtained single-shot measurement results Eq. \ref{eq:singleshotresult}. We acquire three independent data sets, each consisting of 6700 shots in total, for the cases where the measurements take place i) after initialization, ii) after the gates acting on $c$ and $h$, without the SWAP gate and iii) after the SWAP gate between qubits $h$ and $e$. Expectation values $\langle \hat{B}^{\alpha}\rangle$ are computed for all three data sets, by varying $\alpha$. Changes $\delta\langle \hat{B}^{\alpha}\rangle$ with respect to $\alpha$ are then computed for both the cases with and without SWAP gate, with respect to the expectation values computed for the initial state. The results are shown in Fig. \ref{fig:result1}.  Estimates for the statistical error are computed via non-parametric bootstrapping: Artificial event rates of detecting $\{\ket{0_c0_h},\ket{0_c1_h},\ket{1_c0_h},\ket{1_c1_h}\}$ are generated by drawing event numbers from a multinomial distribution, governed by the measured event rates. The artificial rates are used for computing expectation values $\langle \hat{B}^{\alpha}\rangle$, which are used in turn to compute a 1$\sigma$ error channel.\\ 
For the case without SWAP gate, we observe $\delta\langle \hat{B}^{\alpha}\rangle \geq 0$ for the entire range of $\alpha$ values, which indicates a unitary evolution of qubits $c$ and $h$. In contrast, for the case with SWAP gate, we observe $\delta\langle \hat{B}^{\alpha}\rangle \leq 0$ for values of  $\alpha$ below 0.5090(75). This shows a clear violation of the global passivity inequality Eq. \ref{eq:gp3ineq}. Note that the microscopic form of the second law ($\alpha=1$) Eq. \ref{eq:gp1ineq} provides $\delta\langle \hat{B}\rangle \geq 0$, which confirms that the framework of global passivity provides an increased sensitivity for experimental verification of heat leaks.

\begin{figure}[h]
    \centering
    \includegraphics[width=0.9\columnwidth,trim={0 0 21cm 0 },clip]{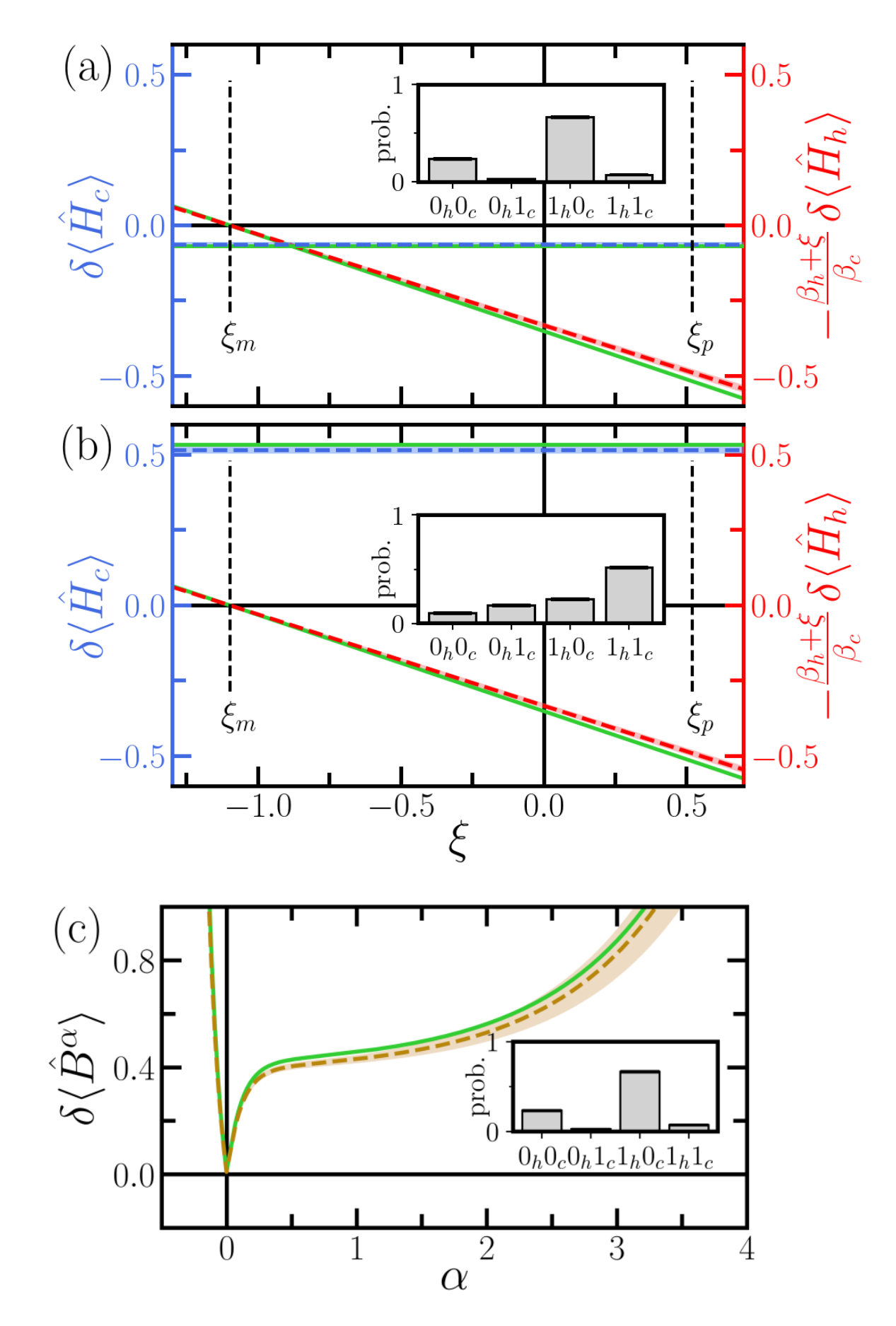}
    \caption{For the second protocol, depicted in Fig. \ref{fig:circuits}(b), we compare the energy change in the cold bath (blue) to some function of the parameter $\xi$ times the energy change in the hot bath (red), for the cases \textbf{(a)} with final SWAP gate, and \textbf{(b)} without final SWAP gate. We find a violation for $-\beta_h\le \xi \le -0.880(1)$ in the case with SWAP gate. Experimental data (blue/red, dotted) is compared to theoretical expectation values (green). The binary probabilities for system qubits $h,c$, from which the data was calculated, are shown as an inset. 3200 shots were used. Panel \textbf{(c)} shows that we observe no violation of the global passivity inequality (Eq. \ref{eq:gp3ineq}) in this case, for the data including final SWAP gate.}
    \label{fig:result2}
\end{figure}

\textit{Heat leak detection via passivity deformation - } A slightly modified protocol serves for detecting the heat leak via the passivity deformation approach. For this case, we choose $\beta_c =\ $1.627(7), $\beta_h =\ $1.099(8) and $\beta_e =\ $2.232(5). As depicted in Fig. \ref{fig:circuits}(b), the joint local qubit rotations are replaced by a rotation about 2.5~rad of qubit $h$ only, described by
\begin{equation}
    \hat{U}_y = \exp\left(-i\;2.5\;\hat{\sigma}_y^{(h)}\right),
\end{equation}
The phase gate between qubit $c$ and $h$ is replaced by a SWAP gate. 
Similarly to the global passivity test, qubits $c$ and $h$ are separated after the unitary evolution, then qubits $h$ and $e$ are merged to the LIZ, where they can undergo an optional SWAP gate.\\
We chose the Hamiltonian of the hot qubit $\hat{H}_h=\mathbb{1}_c \otimes \ket{1_h}\bra{1_h}$ to be the deformation operator $\hat{A}$ (cf. Eq. \ref{eq:PD0}). Both operators $\hat{B}$ and $\hat{B}'$ are diagonal, with eigenvalues
\begin{eqnarray}
\text{eig}_{\uparrow}(\hat{B})&=&\{0,\beta_h,\beta_c,\beta_h+\beta_c\} \nonumber \\
\text{eig}(\hat{B'})&=&\{0,\beta_h+\xi,\beta_c,\beta_h+\beta_c+\xi\}.
\end{eqnarray}
Note that the eigenvalues of $\hat{B}$ are sorted, while the eigenvalues of $\hat{B}'$ are not. Condition Eq. \ref{eq:PD1} requires the eigenvalues of $\hat{B}'$ to have the same sorting as for $\hat{B}$, which leads to
\begin{equation}
\xi_m =-\beta_h\le \xi \le \beta_c-\beta_h=\xi_p 
\end{equation}
The passivity deformation inequality Eq. \ref{eq:PD2} then yields
\begin{equation}
   \delta\langle\hat{H}_c\rangle \ge -\frac{ \beta_h+\xi}{\beta_c}\delta\langle\hat{H}_h\rangle  \quad \forall\ \xi_m  \le \xi \le \xi_p
   \label{eq:deformexamp}
\end{equation}
The left-hand side of this inequality, computed from the measurement data, is shown for varying $\xi$ in Fig. \ref{fig:result2}. For the case with SWAP, we observe a clear violation of Eq. \ref{eq:deformexamp} for $\xi<-0.880(1)$, which is about 5.3 standard deviations above the bound $\xi_m$. From Fig. \ref{fig:result2} c), we see that global passivity fails to detect the heat leak for this scenario, as $\delta\langle \hat{B}^{\alpha}\rangle \geq 0$ for any $\alpha$. This demonstrates that passivity deformation based inequalities yield increased sensitivity to heat leaks as compared to global passivity.

\textit{Conclusion - } We have used trapped ions to demonstrate the relevance of passivity-based inequalities for detecting controllable heat leaks, i.e. the presence of measurable interactions with the environment. While a formulation of a diagnostics scheme based on these ideas requires further study, this experiment shows that passivity based constraints are experimentally relevant and that they are more sensitive to heat leaks as compared to the second law of thermodynamics.\\
Future work will aim on using periodically repeating protocols to amplify the effect of a heat leak and therefore increase the detection sensitivity, in order to detect genuine heat leaks rather than artificially introduced environments in quantum devices.

\begin{acknowledgments}
	FSK and UGP acknowledge funding from DFG with research unit \textit{Thermal Machines in the Quantum World} (FOR 2724), from the EUH2020-FETFLAG-2018-03 under Grant Agreement no.820495 and by the Germany ministry of science and education (BMBF) within IQuAn. RU is grateful for support from Israel Science Foundation (Grant No. 2556/20).
\end{acknowledgments}	

\bibliography{lit_etal}
	
\end{document}